\documentclass[aps,pra,twocolumn, showpacs]{revtex4}
\usepackage{graphicx}
\usepackage{dcolumn}
\usepackage{bm}

\begin{document}
\preprint{preprint}
\title{Optical Bistability and Collective Behavior of Atoms trapped in a High-Q Ring Cavity}
\author{Th. Els\"{a}sser}
\author{B. Nagorny}
\author{A. Hemmerich}
\affiliation{Institut f\"{u}r Laser--Physik, Universit\"{a}t Hamburg,
Luruper Chaussee 149, D--22761 Hamburg, Germany}

\date{\today}

\begin{abstract}
We study the collective motion of atoms confined in an optical
lattice operating inside a high finesse ring cavity. A simplified
theoretical model for the dynamics of the system is developed upon
the assumption of adiabaticity of the atomic motion. We show that
in a regime where the light shift per photon times the number of
atoms exceeds the line width of the cavity resonance, the
otherwise tiny retro-action of the atoms upon the light field
becomes a significant feature of the system, giving rise to
dispersive optical bistability of the intra--cavity field. A
solution of the complete set of classical equations of motion
confirms these finding, however additional non-adiabatic phenomena
are predicted, as for example self--induced radial breathing oscillations. We
compare these results with experiments involving laser--cooled
$^{85}$Rb atoms trapped in an optical lattice inside a ring cavity
with a finesse of $1.8 \times 10^5$. Temperature measurements
conducted for moderate values of the atom--cavity interaction
demonstrate that intensity--noise induced heating is kept at a very
low level, a prerequisite for our further experiments. When we
operate at large values of the atom--cavity interaction we observe
bistability and breathing oscillations in excellent agreement with
our theoretical predictions.
\end{abstract}

\pacs{32.80.Pj, 42.50.Vk, 42.62.Fi, 42.50.-p}

\maketitle

\section{Introduction}
Atoms regularly spaced in optical lattices are a widely studied
model system of quantum optics \cite{Gri:00} which receives
growing attention also from other areas of physics ranging from
the fields of condensed matter \cite{Jak:00} to quantum
information processing \cite{Cal:00}. Magneto--optic trapping
techniques typically provide optical lattices with an average
inter--atomic distance of several microns, where dipole--dipole or
spin interactions are negligible. While in this case the atomic
dynamics is well understood it appears to be of limited relevance
for most conceivable applications. Schemes that provide mutual
interactions among the atoms should significantly increase the
usefulness of optical lattices. Recently, the formation of optical
lattices inside optical resonators has become a subject of
extensive research \cite{Nag1:03, Nag2:03,Kru1:03, Kru2:03},
because cavity--mediated long range interactions arise, possibly
useful for new quantum computing schemes \cite{Hem:99} or novel
laser cooling methods \cite{Doh:97, Hor:97, Hec:98, Cha:03,
Els:03} which do not rely on cyclic spontaneous emission and thus
apply to a wider class of species without degradation to be
expected at high densities.

The key to exploiting these new concepts is a profound
understanding of the interaction of trapped particles with
intra--cavity light fields. For single atoms interacting with only
a few photons in cavities with a small mode volume well below
10$^{-4}$\,mm$^3$ studies have been successfully performed within
the last few years \cite{Pin:00, Hoo:00}. However, in these
experiments only a few atoms can be trapped for no more than a few
10\,ms due to heating caused by intensity fluctuations of the
intra--cavity field. More recently it has been pointed out, that
high finesse cavities with large mode volumes exceeding 1\,mm$^3$
may be utilized to confine large atomic ensembles with the
advantage of far longer trapping times. In this case the high
frequency noise components of the intra--cavity light field are
significantly suppressed owing to the long lifetime of the photons
\cite{Nag1:03}. Although operation at large detunings from atomic
resonance is required in order to prevent undesirable spontaneous
photons, for sufficiently high values of the cavity finesse the
strong coupling regime can be reached. In this case the collective
interaction strength given by the light shift per photon times the
number of trapped particles exceeds the resonance line width of
the cavity, and thus the otherwise tiny retro--action of the atoms
upon the light field becomes a significant feature of the system.

In this paper we experimentally and theoretically explore the
motion of atoms trapped in an optical lattice formed inside a high
finesse ring--cavity with a large mode volume. We discuss two
different regimes of operation characterized by the strength of
the collective atom--cavity coupling. Our observations for moderate
coupling strengths show that the experimental challenges
introduced by the small cavity resonance bandwidth can be handled
and a stable lattice can be produced. At large atom--cavity
couplings where the backaction of the atoms upon the lattice
potential becomes significant, we find non--linear dynamics of the
intra--cavity field and a collective character of the atomic
motion. In this regime phenomena like dispersive optical
bistability and self--induced breathing oscillations arise.
Non--linear dynamics due to optical pumping and saturation has
been observed in low--finesse resonators operating close to an
atomic resonance \cite{Lam:95}. Only recently absorptive optical
bistability was observed in a small--volume high--finesse cavity
\cite{Sau:03}. Our lattice operates far from resonance where
optical pumping and saturation are negligible and the atoms merely
act as a dispersive medium.

The paper is organized as follows. In  Sec. \ref{splitting} we
present a simple physical picture of the atom--cavity coupling in
terms of coherent forward and backward Rayleigh scattering. In
Sec. \ref{model} we briefly discuss the complete set of
semiclassical equations of motion of the system. These equations
are difficult to solve leading us to develop a simplified but
surprisingly accurate model of the complex system dynamics based
upon the assumption that the atomic sample adiabatically adjusts
to the potential position and depth. A steady--state analysis for
the phase and amplitude of the intra--cavity field predicts
dispersive bistability. In Sec. \ref{expsetup} we discuss the
characteristics of our experimental setup. Both cavity modes are
externally fed by adjustable amounts of power from a single--mode
laser diode. A fast servo--lock keeps the laser frequency in
resonance with one of the travelling wave modes of the cavity. The
properties of this lock and the implications for parametric
heating and expected trapping times are discussed in Sec.
\ref{stabilization}. Noise measurements of the light transmitted
through the cavity together with lifetime and temperature
measurements of the trapped atomic ensemble show that undesired
intra--cavity intensity noise can be maintained at extremely low
values despite of the small cavity bandwidth. In Sec.
\ref{non--linear} we present experiments in the strong coupling
regime. For asymmetric pumping of the cavity the unlocked
intra--cavity field exhibits distinctive non--linear behavior
resulting in bistability of the intra--cavity field. We find an
excellent agreement with the predictions of the adiabatic model of
Sec. \ref{model}. In Sec. \ref{non--adiabatic} we explore
non--adiabatic aspects of the atomic motion. Observations of
radial breathing oscillations are discussed and a numerical
simulation based on solving the full set of equations of motion is
presented.

\section{Mode Splitting}
\label{splitting} 
\begin{figure}
\includegraphics[width=4.5 cm]{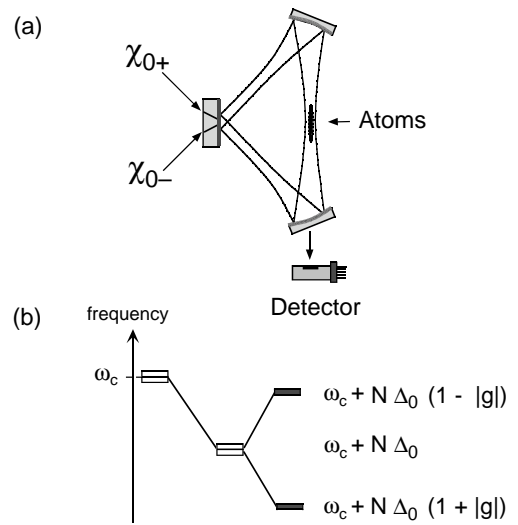}
\caption{ \label{setup}a) Sketch of the experimental setup.
Adjustable laser powers  $\chi_{0\pm}$ are coupled into the
two counterpropagating travelling wave modes of a ring resonator.
The power transmitted through the cavity can be detected by a
photo diode (PD). b) Mode splitting
mechanism (explained in the text).}
\end{figure}

The interaction of the atomic sample with the
intra--cavity light field can  be understood in terms of coherent
Rayleigh scattering. Consider the two degenerate, mutually
counterpropagating, travelling wave modes of a ring resonator
(discriminated by indices ($\pm$) in the following) with a common
resonance frequency $\omega_c$, as depicted in Fig. \ref{setup}a).
Let us add atoms into the common mode volume with a resonance
frequency $\omega_a >> \omega_c$ such that their interaction with
the intra--cavity fields has entirely dispersive character. If the
atomic sample is homogeneously distributed it will merely give
rise to Rayleigh scattering in the forward directions described by
a constant index of refraction $n_{\pm} = 1 + N\, \Delta_0 /
\omega_c$, where $\Delta_0$ is the light shift per photon and $N$
is the number of atoms. This causes a common frequency shift  $N\,
\Delta_0$ for both modes [Fig. \ref{setup}b)]. Assume that both
modes are externally coupled and an optical lattice is formed
inside the cavity which tightly traps the atoms in the Lamb--Dicke
regime, where the atomic distribution in each lattice site is
confined to a fraction of the optical wavelength. In this case
additional backscattering arises, which couples the two
counterpropagating modes and lifts their frequency degeneracy
giving rise to a frequency splitting $2 N \Delta_0 \, |g| $. The
degree of atomic localization is measured by the parameter $g
\equiv \frac{1}{N} \sum_{\nu = 1}^{N} e^{-i2kz_{\nu} -
2(x_{\nu}^2+y_{\nu}^2)/w_0^2}$, where k is the wave number,
$x_{\nu}$, $y_{\nu}$ and $z_{\nu}$ denote the atomic position
coordinates, and $w_0$ is the mode radius. If there is no
statistical correlation between the axial and radial coordinates
and the spatial distributions near each lattice site follow a
Gaussian, we may write $g =   \left(1+4 \left(\sigma_r/ w_0
\right)^2 \right)^{-1} exp(-2k^2 \sigma_z^2) \, exp(-2ik z_{cm})$,
i.e., the complex phase scales with the axial center of mass
coordinate $z_{cm}\equiv \frac{1}{N} \sum_{\nu = 1}^{N} z_{\nu}$
and the modulus is a measure for the axial ($\sigma_z$)
 and radial ($\sigma_r$) spread of the atomic sample. Perfect localization
corresponds to $|g|=1$, whereas a homogeneous distribution is
described by $|g|=0$.

If the interaction strength $N\, \Delta_0$ exceeds the cavity
resonance bandwidth, the eigenmodes of the cavity acquire
orthogonal standing wave geometries. One of them supports the
optical lattice with atoms localized at the antinodes, i.e, the
interaction is maximized and the corresponding frequency shift
$\omega_c + N \Delta_0 \, \left(1 + |g| \right)$ exceeds that
found for the travelling wave eigenmodes in presence of a
homogeneous atomic sample. The second eigenmode is not externally
coupled and its nodes coincide with the atomic center of mass
positions thus minimizing interaction. This leads to a reduced
shift $\omega_c + N \Delta_0 \, \left(1 - |g| \right)$
(Fig.\ref{setup}b). The existence of a second empty eigenmode,
blue--detuned with respect to the optical lattice, can be utilized
for implementation of a sideband cooling scheme as described in
Ref. \cite{Els:03}.

It is instructive to consider the change of the indices of
refraction $n_{\pm}$ experienced by the travelling wave modes, due
to back--scattering. According to the considerations of Ref.
\cite{Wei:98} (Eq. 7), to first scattering order, the modified
refractive indices are given by $n_{\pm} = 1 + N\, \frac{\Delta_0
}{\omega_c} (1 + |g| |\frac{\alpha_{\mp}}{\alpha_{\pm}}|)$. Note
that $n_{+}$ and $n_{-}$ may differ, if the field amplitudes of
the travelling waves $\alpha_{\pm}$ are not equal, e.g., for
asymmetric external pumping of the cavity. This gives rise to
interesting non--linear dynamics, if back--scattering becomes
relevant. A small external pumping asymmetry can yield a relative
change of the refractive indices and a corresponding spatial phase
shift of the optical standing wave potential. As the laser
frequency is kept actively in resonance with say the ($+$)--wave,
the ($-$)--wave is tuned slightly out of resonance and the
potential well depth decreases. This in turn decreases the degree
of localization $|g|$ which reduces the phase shift. As we will
see in the following, in a certain parameter range this dynamics
provides more than one steady--state, thus giving rise to
bistability phenomena.

\section{Adiabatic Model}
\label{model} We restrict ourselves to the case where the cavity
modes support coherent fields with large mean photon numbers
interacting with large thermal atomic samples at sufficiently high
temperatures such that a classical description of the optical
intra--cavity fields and the atomic position and momentum variables
is appropriate. Moreover, we assume a large detuning of the pump
frequency from the atomic resonance and hence negligible
saturation. In this case according to Ref. \cite{Gan:00} the
following equations describe the time evolution of the complex
fields $\alpha_+$ and $\alpha_-$ for the counterpropagating
travelling wave modes scaled to the field per photon

\begin{equation}
\label{ritschfield}
\begin{array}{ccc}
\frac{d}{dt} \left( \begin{array}{ccc} \alpha_+ \\ \alpha_- \\
\end{array} \right) = \bf{M} \left( \begin{array}{ccc} \alpha_+ \\
\alpha_- \\ \end{array} \right) + \gamma_0 \left(
\begin{array}{ccc} \eta_+ \\ \eta_- \\ \end{array} \right)
\\
\\
\bf{M} \equiv \left( \begin{array}{cc}
i \left( \delta_c - N \Delta_0 g_{r} \right) - \gamma_c & - i N \Delta_0 \, g \\
- i N \Delta_0 \, g^* &  i \left( \delta_c - N \Delta_0  g_{r} \right) - \gamma_c \\
\end{array} \right)
\\
\\
g_{r} \equiv \frac{1}{N} \sum_{\nu = 1}^{N}
e^{-2(x_{\nu}^2+y_{\nu}^2)/w_0^2}
\end{array}
\end{equation}

Here $\eta_+$ and $\eta_-$ are the complex amplitudes of the
incoupled light fields,  and $\delta_c$ denotes the detuning of
the incoupled frequency from the resonance frequency $\omega_c$ of
the empty cavity. The radial bunching parameter $g_r$ accounts for
the radial position dependence of the atom--light coupling. The
diagonal elements of the matrix $\bf{M}$ comprise the forward
scattering terms $ N \Delta_0  g_{r}$ which only depend on the
degree of radial bunching while the off--diagonal terms which act
to couple $\alpha_+$ and $\alpha_-$ are due to backward scattering
and thus involve the localization parameter $g$.

The equations for the momenta $\vec{p}_\nu  \equiv (p_{x,\nu}$,
$p_{y,\nu}$, $p_{z,\nu})$ and the positions $\vec{x}_\nu \equiv
(x_\nu, y_\nu, z_\nu)$ read

\begin{equation}
\label{coordinates} \begin{array}{ccc}
\frac{d}{dt} \vec{p}_\nu  =
\hbar \Delta_0 \left( \vec{\nabla} \, \, {|\alpha_+ e^{-i kz} +
\alpha_- e^{i kz}|}^2 e^{- 2 \frac{(x^2+y^2)}{w_0^2}}
\right)\big|_{\vec{x}=\vec{x_\nu}} \\
\\
\frac{d}{dt} \vec{x}_\nu = \frac{1}{m} \vec{p}_\nu
\end{array}
\end{equation}

In total we are concerned with $6N+2$ coupled first--order
non--linear differential equations for the $3N$ position and
momentum variables, respectively and the two complex field
amplitudes for the counter--propagating cavity modes.

In experiments with small bandwidth cavities the frequency of the
incoupled laser beam needs to by actively controlled in order to
maintain some resonance condition, which allows to couple light
into the cavity at all. In our implementation we servo lock the
laser frequency to one of the counterpropagating travelling waves
(say the (+)--mode) with a technique discussed in detail in
\ref{stabilization}. Here we merely account for the fact that this
servo acts to maintain the complex field amplitude $\alpha_{+}$ at
a constant value, i.e., $\frac{d}{dt} \alpha_+ \equiv 0$. Hence,
we may combine the two field equations (\ref{ritschfield}) into a
single equation for the unlocked mode $\alpha_{-}$

\begin{equation}
\label{fieldnew} \frac{d}{dt} \alpha_- = i \frac{\Delta_0
N}{\varepsilon_+} g \alpha_-^2 - \alpha_- \gamma_c - i \Delta_0 N
g^* \varepsilon _+ + \varepsilon _- \gamma_c \,\,\, .
\end{equation}

using the abbreviations $\varepsilon _\pm \equiv
\frac{\gamma_0}{\gamma_c} \eta_\pm$ for the steady state
intra--cavity fields in absence of atoms. We may choose
$\varepsilon _\pm$ to be real without loss of generality, thus
fixing the spatial phase of the intra--cavity standing wave for
the case when no atoms are present.

For numerical processing and comparison with experiments it is
useful to work with unit--free quantities. We hence define scaled
fields $a \equiv \alpha_-/\sqrt{I_0}$ and $\sqrt{\chi_{0\pm}}
\equiv \varepsilon_\pm/\sqrt{I_0}$ with $I_0 \equiv  |\varepsilon
_+|^2 + |\varepsilon _-|^2$ being the sum of the steady state
intensities of each travelling wave mode in absence of atoms.
Using the scaled time $\tau \equiv \gamma_c t$ and the scaled
interaction strength $U \equiv \Delta_0/\gamma_c$ we obtain

\begin{equation}
\label{fieldscaled} \frac{d}{d \tau} a = i \frac{U
N}{\sqrt{\chi_0+}} |g| \hat{g} a^2 - a  - i U N |g| \hat{g}^*
\sqrt{\chi_{0+}} + \sqrt{\chi_{0-}} \,\,\, .
\end{equation}

In order to incorporate the equations of motion for the atomic
ensemble into this equation we seek expressions for the complex
phase $\hat{g}=g/|g|$ of the localization parameter $g$, which is
connected to the atomic center--of--mass coordinate, and the modulus
$|g|$ which represents the spatial spread of the atomic sample in
each potential well. Therefore, let us consider atoms tightly
confined in the optical lattice, such that the time scale of the
axial motion is much shorter than the photon life time
$(1/2\gamma_c)^{-1}$ which represents a lower bound for the time
scale relevant for changes of the intra--cavity fields. In this
case it is reasonable to expect, that the center--of--mass in each
potential well of the optical standing wave adiabatically follows
the potential minimum, which formally is expressed by the relation
$\hat{g}^*=a/|a|$. More specifically, we assume that the atomic
distribution in the lattice can be described by a sum of Gaussians
which are centered at the instantaneous antinodes of the lattice.
To evaluate $|g|$ we take into account that there is no
statistical correlation between the axial and radial coordinates
writing $|g| \approx \langle e^{-i 2 kz} \rangle \langle e^{-2
r^2/w_0^2} \rangle$, where the brackets denote a Gaussian average.
These expressions are readily calculated to be $\langle e^{-i 2 k
z} \rangle = e^{-2 k^2 \sigma_z^2}$ and $\langle e^{-2 r^2/w_0^2}
\rangle = 1/(1+4\sigma_r^2/w_0^2)$. In order to determine the time
evolution of $\sigma_r$ and $\sigma_z$ we assume that the thermal
atomic sample adjusts adiabatically to the potential well depth
keeping the Boltzmann factors $\xi_{ax} \equiv k_B
T_{ax}/\omega_{v,ax}$ and $\xi_{rad} \equiv k_B
T_{rad}/\omega_{v,rad}$ constant, where $k_B$ is the Boltzmann
constant and $T_{ax}$ and $T_{rad}$ are the axial and radial
temperatures of the sample and $\omega_{v,ax}$ and
$\omega_{v,rad}$ denote the axial and radial trap frequencies.
By harmonically approximating the potential in axial and radial directions we
find $2 \sigma_z^2 /k^2 = k_B T(t) / V_{0,ax} = \eta_{ax} \sqrt{a(0) / a(t)}$ and
$4 \sigma_r^2 / w_0^2 = \eta_{rad} (\sqrt{\chi_{0+}} +
|a(0)|)/(\sqrt{\chi_{0+}} + |a(t)|)$,
with $\eta_{ax}$ and $\eta_{rad}$ being the ratio between the thermal
energy and the potential depth at t=0 for the axial and radial directions,
respectively. With these approximations we are in the position to
state a single equation for the complex unstabilized electric
field amplitude:

\begin{eqnarray}
\label{adiabatic} \frac{d}{d \tau} a \, = \, i \frac{U
N}{\sqrt{\chi_{0+}}} \, L(a) \, |a| a \, - \, a  \, +   {}
\nonumber \\
+ \, \sqrt{\chi_{0-}} \, - \, i U N \, \sqrt{\chi_{0+}} \, L(a) \,
\frac{a}{|a|} \, \, ,
\nonumber \\
L(a) \equiv e^{-   \eta_{ax} \sqrt{\frac{|a_{0}|}{|a|}}   } \, \,
\frac{1}{1 + \eta_{rad} \, \frac{\sqrt{\chi_{0+}}\, + \,
|a_{0}|}{\sqrt{\chi_{0+}} \, + \, |a|}}
\end{eqnarray}

In this equation all the parameters can be easily obtained by
measurements and a numerical integration yields simulations of the
intra--cavity field, that can be compared to experimental data. In
Sec. \ref{non--linear} we will show that despite its simplicity the
adiabatic model presented here reproduces our observations very
accurately.

We may obtain additional physical insight into the dynamical
properties of Eq. (\ref{adiabatic}) by representing the complex
field $a$ as $a=|a|e^{i\phi}$ with amplitude $|a|$ and phase
$\phi$. Multiplying Eq. (\ref{adiabatic}) by $a^*/|a|$ results in
separate differential equations for $|a|$ and $\phi$:

\begin{eqnarray}
\label{phase} |a| \frac{d}{d\tau} \phi = U N L(a) (
\frac{|a|^2}{\sqrt{\chi_{0+}}} - \sqrt{\chi_{0+}}) -
\sqrt{\chi_{0-}}\sin{\phi} \\
\label{amplitude} \frac{d}{d\tau} |a| = \sqrt{\chi_{0-}}\cos{\phi} - |a|
\end{eqnarray}

Eq. (\ref{amplitude}) implicates that the amplitude $|a|$ adjusts
exponentially to $\sqrt{\chi_{0-}} \cos{\phi}$ determined by the
instantaneous value of $\phi$. This happens at the fastest
time--scale available inside the cavity, the decay time of the
intra--cavity field. Hence, any changes of the intra--cavity field
on slower time--scales are governed by the evolution of $\phi$.
This lets us adiabatically eliminate Eq.~(\ref{amplitude}) by
inserting $|a | = \sqrt{\chi_{0-}} \cos{\phi}$ into
Eq.~(\ref{phase}), which leads to

\begin{eqnarray}
\label{phasess}\frac{d}{d\tau} \phi = \frac{ U N }{\sqrt{\chi_{0-} \, \chi_{0+}}}
\widetilde{L}(\phi)
\left(\chi_{0-} \cos{\phi} - \frac{\chi_{0+}}{\cos{\phi}} \right)
\\ \nonumber \\
- \sqrt{\chi_{0-}}\tan{\phi} \nonumber
\\ \nonumber \\
\widetilde{L}(\phi) \equiv \frac{\exp{\left(-\xi_{ax}
\frac{\omega_R}{\omega_V}\sqrt{\frac{8}{\sqrt{\chi_{0+}\chi_{0-}}
\cos{\phi}}}\right)}}{1 + \frac{8 \xi_{rad} \omega_R}{k w_0
\omega_V (\sqrt{\chi_{0+}}+\sqrt{\chi_{0-}} \cos{\phi})}}\nonumber
\end{eqnarray}

The localization factor $\widetilde{L}(\phi)$ is rescaled in terms
of quantities which remain constant during the time evolution of
the system, i.e. the axial  and radial Boltzmann factors
$\xi_{ax}$ and $\xi_{rad}$, the recoil frequency $\omega_R$ the
mode radius $w_0$ and an effective axial vibrational frequency
which is a measure of the total power directed to the cavity. More
precisely, $\omega_V$ is the axial vibrational frequency
corresponding to the optical potential that would arise for
symmetric pumping with no atoms inside the cavity, i.e. if
$\alpha_{+} = \alpha_{-} = \varepsilon_{+} = \varepsilon_{-}$.

\begin{figure}
\includegraphics[width=\columnwidth]{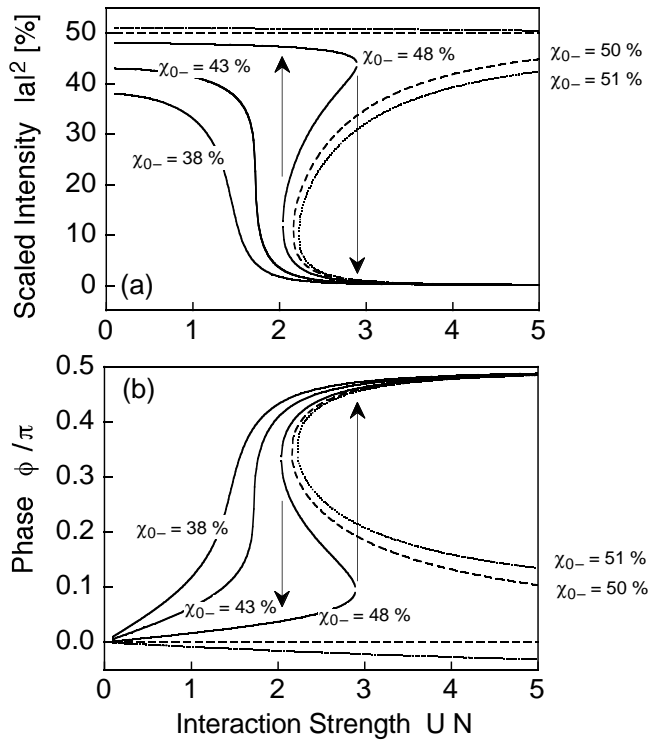}
\caption{Steady state solutions of Equations
(\ref{phase},\ref{amplitude}) for the intensity (a) and the phase
(b) of the unlocked intra--cavity field for $\chi_{0-} = 51\%,
50\%, 48\%, 43\%$ and $38\%$. For $\chi_{0-}<43\%$ only one stable
solution exists, while for larger values bistability occurs for $U
N$ larger than some value around two. The arrows indicate for the
$48\%$--trace, where a jump from one to the other stable solution
can occur, if the interaction parameter is tuned.}
\label{bistability}
\end{figure}

The dynamical properties of Eq.~(\ref{adiabatic}) can be analyzed
in terms of its steady state solutions. The  steady state values
of $\phi$ are obtained as the zeros of the the right hand side of
Eq.~(\ref{phasess}) while the corresponding values of $|a|$ follow
directly from Eq.~(\ref{amplitude}). It is particularly
instructive to plot these steady state values versus the
interaction strength $U\,N$, which can be tuned experimentally. In
Fig.~\ref{bistability} this is shown for different degrees of
pumping asymmetry, i.e., $\chi_{0-} = 51\%, 50\%, 48\%, 43\%$ and
$38\%$. In this plot we use our experimental data for $\xi_{ax},
\xi_{rad}, \omega_V, k$ and $w_0$. For values of $\chi_{0-} <
43\%$ only one stable solution exists and the amplitude and phase
of the intra--cavity field $a$ are well defined. The situation
changes with increasing $\chi_{0-}$. For higher values, e.g.
$\chi_{0-} = 48\%$ a strong bistability occurs for an interaction
strength of $2 \lesssim U\,N \lesssim 3$. The upper and lower
branch with a positive (negative) slope for the phase (amplitude)
are stable, whereas the middle branch is unstable. Experimentally,
the interaction strength is readily tuned by a change of the
particle number $N$. If one increases $UN$ starting from low
values the system will follow the upper branch of the
$48\%$--trace in Fig. \ref{bistability}a) until the turning point
at $U\,N \approx 3$ is reached, where the intensity suddenly drops
to almost zero. On the other hand, when we reduce the number of
particles starting from high values, the intensity abruptly jumps
from small to high intensity around $U\,N \approx 2$. This
hysteresis feature is characteristic for bistable systems. If we
approach symmetric pumping (i.e. $\chi_{0-}$ approaches $50 \%$
the upper bound of the bistability range (the region on the
x--axis between the jumps) moves further out towards infinity
whereas the lower bound only slightly increases above two. For
$\chi_{0-} \geq 50 \%$ the upper bound of the bistability range
equals infinity, i.e., the low intensity branch cannot be reached
by tuning $U\,N$. In this case a stable optical lattice can be
obtained at any value of the interaction strength.

In our analysis we have neglected non--adiabatic aspects of the
atomic motion although this is only justified for the axial
degrees of freedom which are well confined. Motion along the
radial directions occurs on a far slower time scale, and the
adiabatic approximation does not hold. A typical non--adiabatic
reaction of the radial motion to sudden changes of the potential
depth is the excitation of breathing oscillations. Such
oscillations would come along with a small oscillatory change of
the effective interaction strength. If the system operates deeply
inside the bistability regime we may immediately infer from
Fig.~\ref{bistability} that small changes of $UN$ with an
amplitude well below the extension of the bistability range do not
significantly change the intra--cavity intensity. The most drastic
reaction of the intensity to a change of $UN$ occurs, if the
system is close to the frontier between stable and bistable
operation, as in the $43\%$ trace of Fig.~\ref{bistability}. In
this case we in fact observe oscillations of the intensity which
are not predicted by our adiabatic model and which are subject of
Sec. \ref{non--linear}.

\begin{figure}
\includegraphics[width=\columnwidth]{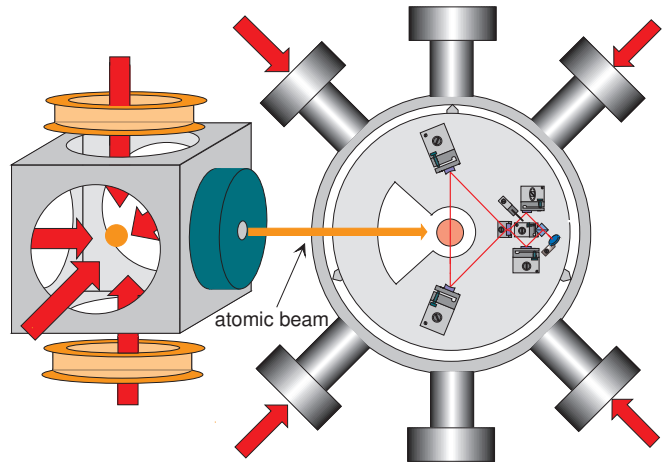}
\caption{Sketch of the double--MOT system with the resonator placed
in the vacuum chamber.} \label{doublemot}
\end{figure}

\section{Experimental Setup}
\label{expsetup} Preparation of ultracold $^{85}$Rb atoms is
established with a standard double--MOT (magneto--optical trap)
setup as depicted in Fig.~\ref{doublemot}. In a first vacuum
chamber a conventional 3D--MOT collects atoms from a Rb dispenser
source. One of the retro--reflecting mirrors is provided with a
hole, such that an atomic beam is extracted with a flux of $5
\cdot 10^8$\,s$^{-1}$ and a mean velocity of 10\,m/s. This beam
loads a second magneto--optical trap placed in the main UHV chamber
with a pressure below $2\cdot 10^{-10}$\,mbar. Here we typically
trap $5\cdot10^8$ atoms at a temperature of 100\,$\mu$K. The
implementation of the MOT coils inside the vacuum allows us to
switch the magnetic field very rapidly. Contamination emerging via
heating of these coils limits the lifetime of the trap to about
2\,s.

The experimental implementation of the resonator is also sketched
in Fig.~\ref{doublemot}. In a triangular setup two curved high
reflecting mirrors (1.5\,ppm transmission) and the flat incoupling
mirror (23\,ppm transmission) form the high finesse ring cavity
with a round trip path length of 97\,mm and a free spectral range
of 3.1\,GHz. The beam waists $w_0$ ($1/e^2$ mode radius) in
sagittal and vertical directions are 134\,$\mu$m and 129\,$\mu$m,
respectively. The beam splitting optics is included in the vacuum
to keep the path lengths between the splitter and the incoupling
mirror as short as possible. This is favorable because the path
length difference influences the spatial phase of the
intra--cavity standing wave.

For the trapping light we use an extended cavity grating
stabilized diode laser with variable detuning relative to the D2
transition of $^{85}$Rb. An acoustooptical modulator (AOM) serves
as a fast switch and an intensity regulator for the lattice. We
analyze the reflected and transmitted light with three avalanche
photo diodes.

\section{Stabilization, parametric heating, trapping time}
\label{stabilization} In order to stabilize the laser diode
emission to the cavity resonance, we use a frequency modulation
technique (Pound--Drever--Hall) with a servo bandwidth of several
MHz. Similarly as in Ref. \cite{Sch:01} the fast branch of the
feedback control directly applies the PDH--error signal to the
injection current of the diode after passing a loop filter, which
compensates for frequency--dependent phase shifts on the diode
laser chip. With the laser locked to the resonator we can measure
the photon lifetime $1/2 \gamma_c$\,=\,9.3\,$\mu$s corresponding
to a finesse of the cavity of F\,=\,$1.8 \cdot 10^5$, a cavity
resonance line width of 17\,kHz and a mean scattering loss per
mirror of 3\,ppm.

A sufficiently fast and efficient stabilization is needed to
prevent intensity fluctuations of the intra--cavity field which
lead to exponential heating of the trapped atoms \cite{Sav:97}. In
particular, in the case of a small cavity resonance line width
tiny frequency deviations are readily converted into intensity
fluctuations. Unfortunately, the heating time scales quadratically
with the trap frequency, which typically is on the order of a few
hundred kHz for the axial direction in a standing wave trap
\cite{Sav:97}:

\begin{equation}
\dot{W} = \frac{1}{3}\gamma_a W + \frac{2}{3}\gamma_r W \quad \,
\gamma_{a,r}=\pi^2 \nu_{a,r}^2 S(2 \nu_{a,r})
\end{equation}

\begin{figure}
\includegraphics[width=\columnwidth]{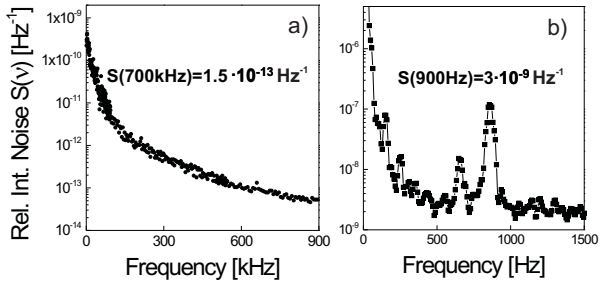}
\caption{Relative intensity noise power spectrum. a) High
frequency part of the spectrum up to 1\,MHz, relevant for axial
heating b) Low--frequency part of the spectrum responsible for
radial heating. The pronounced peak at $\approx$\,850\,Hz
originates from the turbo pump.} \label{noise}
\end{figure}

Here W is the mean kinetic energy and $\nu_{a,r}$ are the trap
frequencies of the atoms in axial and radial direction,
respectively. $S(2\nu)$ denotes the one--sided power spectrum of
the fractional intensity noise. The heating rate involves the
noise intensity at the second harmonic of the vibrational
frequency because parametric excitation is the main source of
heating. Here, we assumed that the atomic sample is thermalized
resulting in the same mean kinetic energy for all dimensions.
Nevertheless, the heating rates for the radial and axial
directions may be significantly different. We have measured the
intensity fluctuations $S(2\nu)$ by analyzing the transmission
through one of the high reflectors. The resulting power spectrum
of the relative intensity noise is shown in Fig. \ref{noise}. For
typical trap parameters $\nu_a$=350\,kHz and $\nu_r$=450\,Hz we
find S(700\,kHz)=$1.5\cdot 10^{-13}$\,Hz$^{-1}$ and
S(900\,Hz)=$3\cdot10^{-9}$\,Hz$^{-1}$ resulting in a heating time
(time for an e--fold temperature increase)
$\tau_h=(1/3\gamma_a+2/3\gamma_r)^{-1}$ exceeding 20\,s. This
indicates that heating due to intensity noise should not be a
limiting factor for experiments with trapping times up to several
seconds. The decrease of the intensity noise with 20\,dB per
decade starting at $\sim$17\,kHz is not surprising, because the
resonator acts as a low pass filter due to the long photon
lifetime. Obviously, the small cavity line width only complicates
the stabilization of the intra--cavity field with respect to low
noise frequencies, while high noise frequencies are strongly
suppressed.

\begin{figure}
\includegraphics[width=7cm]{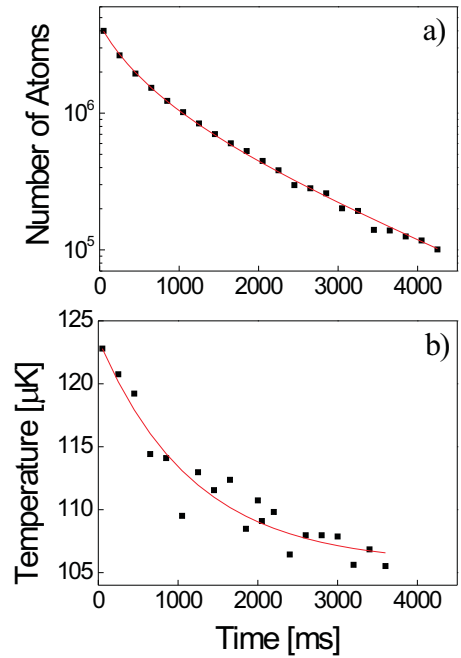}
\caption{a) Temporal evolution of the trap population determined
by fluorescence imaging. The solid line represents a fit to a
model which accounts for two--body losses.
b) Temperature evolution with time. The solid line shows a fit of
Eq.\ref{tempevolution} to the experimental data} \label{temp}
\end{figure}

Our noise analysis promises trapping times on the order of a few
seconds limited merely by background collisions. In order to
confirm this prediction we have tuned the optical lattice 7\,nm
below the D2--line of $^{85}$Rb, where the collective interaction
parameter $U\,N$ is well below unity. In this weak coupling regime
we have conducted life time and temperature measurements. We
couple a power of 60\,$\mu$W into each running wave mode, which is
enhanced by a factor of about 70000 to produce a trap depth of
350\,$\mu$K corresponding to axial and radial frequencies of
350\,kHz and 450\,Hz, respectively.

We typically load $4\cdot10^6$ atoms into the lattice with an
initial temperature of about 100\,$\mu$K. In Fig. \ref{temp}a) we
plot the observed trap population versus time. It shows a
non--exponential two--body decay at the beginning with an
exponential decay time of 1.7\,s representing the main
contribution after 500\,ms. The solid line is a fit according to a
standard trap decay model including a two--body loss term. The
two--body losses are explained by evaporation arising due to the
relatively high density of $9\cdot10^{11}$\,cm$^{-3}$ and the
large elastic scattering cross section of $^{85}$Rb. The expected
reduction of the temperature T(t) is shown in Fig. \ref{temp}b).
According to a simple theoretical model presented in
Ref.~\cite{Nag1:03} the temporal temperature evolution follows

\begin{eqnarray}
\label{tempevolution}
T(t) = T(0) \, \left( 1 - \epsilon \, \frac{\beta \rho_0}{4 \gamma} \left( 1 - e^{- \gamma t} \right)  \right), {}\nonumber \\
\epsilon = \frac{2}{3 k_B T(0)} \overline{W} -1    \, \,
,\\\nonumber
\end{eqnarray}

where $\beta$ is the two--body loss coefficient, $\gamma$ the
exponential decay due to background losses and $\rho_0$ the peak
density in the potential wells. $\overline{W}$ denotes the mean
kinetic energy removed by an evaporated particle. The solid line
of Fig.~\ref{temp}b) shows a fit to the experimental data by means
of Eq.~\ref{tempevolution} with $\epsilon = 0.23$ used as a
fitting parameter, while $\beta$ and $\gamma$ are taken from
Fig.~\ref{temp}a) and $\rho_0$ is measured directly.

According to Fig.~\ref{temp}b) we do not observe heating at least
for the first four seconds of the time evolution that we can
follow before the trapped sample disappears. This lets us
determine a lower bound for the heating time of about 100\,s which
is a factor of 4 larger than the value of $\tau_h$ taken from the
noise measurements. This is not surprising since the temperature
is determined for the radial direction, whereas the main
contribution to heating originates from the axial direction. The
heating rate along the cavity axis is almost two orders of
magnitude higher as compared to radial heating. While during the
first 500\,ms thermalization keeps the axial and radial
temperatures locked, this is not so for later times, where the
density has decreased beyond the collisional regime and radial
heating should slow down significantly.

\section{Strong coupling regime} \label{non--linear}
The strong coupling regime is accessed if the collective
interaction strength $\Delta_0\,N$ exceeds the decay time of the
intra--cavity field $\gamma_c$ ($UN>1$). In order to keep the
spontaneous scattering time of several ms long enough to avoid
significant heating, we operate the lattice at $0.7$\,nm detuning.
This yields a light shift per photon $\Delta_0$=0.091\,s$^{-1}$.
Hence with a few $10^6$ atoms trapped in the lattice we are able
to reach values for $UN$ up to five, well within the strong
coupling regime. In the following experiments we have typically
applied 5\,$\mu$W laser power for each travelling wave mode
yielding a trap depth of about 800\,$\mu$K for symmetric pumping
and in absence of atoms. The corresponding axial vibrational
frequency of 500\,kHz is sufficiently larger than the 17\,kHz
cavity bandwidth, i.e., adiabaticity of the axial motion is a well
justified assumption. A typical experiment proceeds in three
steps. We superimpose the MOT upon the optical lattice for several
seconds, before the MOT light is shut off and the atoms remain
trapped in the lattice. Finally, the lattice is extinguished and
the atoms are given some time to expand ballistically before a
fluorescence image of the sample is taken. From the time of flight
the radial temperature of the sample is determined. Without
ballistic expansion the fluorescence images give information on
the spatial distribution of the trapped atoms. We can continuously
tune the ratio of powers in the two running wave modes from the
case of symmetric pumping $\chi_{0-}=\chi_{0+}$ to one--sided
pumping. We lock the laser frequency to only one of the modes,
i.e. $\chi_{+}$, and keep its phase and amplitude at a constant
value. The power leaking out from the unlocked mode through one of
the high--reflectors is detected and used to determine the scaled
intensity $\chi_-$, which corresponds to $|a|^2$ in the
theoretical model in Sec. \ref{model}.

\begin{figure}
\includegraphics[width=\columnwidth]{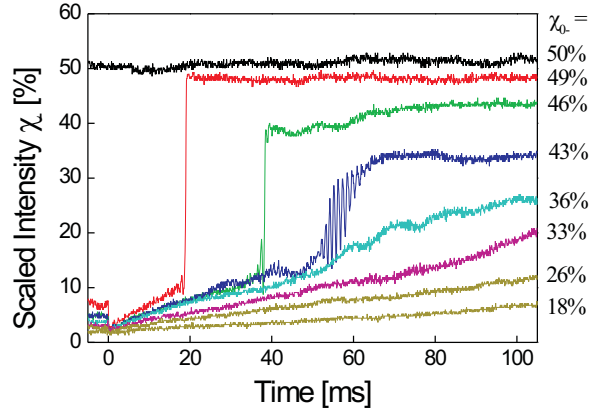}
\caption{Scaled intensity $\chi_-$ for different values of
$\chi_{0-}$. The MOT beams are shut off at $t=0\,ms$.}
\label{intensities}
\end{figure}

\begin{figure}
\includegraphics[width=\columnwidth]{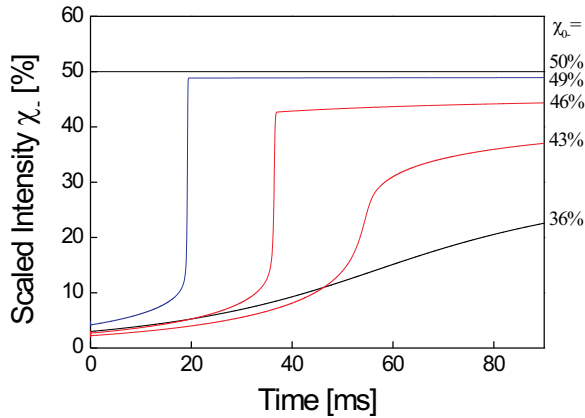}
\caption{Theoretical simulations of the experimental observations of
Fig. \ref{intensities}.} \label{simulation}
\end{figure}

In Fig. \ref{intensities} we show the time evolution of the scaled
intra--cavity intensity $\chi_-$ for different types of pumping.
The MOT is terminated at t = 0 in this figure. For symmetric
pumping ($\chi_{0\pm}$=$50\%$) a stable lattice is formed with
equal intensities in each travelling wave mode irrespective of the
number of atoms inside the cavity. For values of $\chi_{0-}$ below
$50\%$ the situation changes drastically. For an asymmetry of only
$1\%$ in favor to the locked mode the intra--cavity intensity
$\chi_-$ drops to about 10 percent during the MOT phase and to
almost zero after shutting off the MOT beams. Subsequently, a slow
increase is observed before at about 20\,ms the intensity suddenly
jumps back to nearly the value $\chi_{0-}$ (the value expected in
absence of atoms) with a rise time below a few hundred
microseconds. If the power of the unlocked mode is further
reduced, the time duration before the jump occurs increases as
well as the rise time until the jump completely vanishes for
values $\chi_{0-}<40\%$. At a value of ($\chi_{0-}$=$43\%$), where
the jump begins to level off, strong oscillations of roughly
1\,kHz are observed corresponding to twice the radial vibrational
frequency. The traces for asymmetric pumping in
Fig.~\ref{intensities} are observed for $\chi_{0-}$ adjusted to
$49 \%, 46 \%, 43 \%, 36 \%, 33 \%, 26 \%$, and $18 \%$
respectively. The corresponding values of the initial interaction
strength $U~N(t=0) \approx$ 4.48, 4.25, 4.01, 3.54, 3.30, 2.95,
2.48 are carefully determined by measuring the initial particle
number $N(t=0)$ via fluorescence detection. The observed decrease
of $N(t=0)$ with decreasing $\chi_{0-}$ arises because the capture
efficiency decreases with the lattice well depth.

In our experiments the value of the interaction strength
necessarily decreases with time due to trap loss. The
corresponding decrease of $U\,N$ in connection with the
bistability plot of Fig.~\ref{bistability}a) explains the
observations of Fig.~\ref{intensities}. As time proceeds in
Fig.~\ref{intensities} we move from right to left in
Fig.~\ref{bistability}a) starting on one of the low intensity
branches in the lower right corner. Hence, depending on the value
of $\chi_{0-}$ we encounter a sudden or soft increase of
intensity, depending on whether we travel on a curve with bistable
or stable character. We also adjusted values of $\chi_{0-}$ above
$50\%$. In this case $\chi_-$ initially drops to a value close to
$50\%$ independent of the value of $\chi_{0-}$ and gradually
recovers to $\chi_{0-}$. This behavior is understood by similar
arguments based on Fig.~\ref{bistability}.

We have simulated the time evolution observed in
Fig.~\ref{intensities} by means of Eq.~\ref{adiabatic}. The values
of $|a_0|$ are taken directly from the observations of
Fig.~\ref{intensities} a) for $t=0$. The values of $\eta_{ax}=0.5$
and $\eta_{rad}=0.3$ are determined by temperature measurements
with an uncertainty of about 0.1. The difference in the radial and
axial directions originates from different trap depths for both
directions, since the contrast of the interference pattern in
axial direction depends on the degree of pumping asymmetry. The
decrease of $N(t)$ with time is measured and modelled as described
in Sec.~\ref{stabilization}. The theoretical simulations shown in
Fig.~\ref{simulation} reproduce our experimental traces very
nicely. Not only the general behavior of the jump feature, but
also the time of the jump to occur, is accurately matched. For the
initial interaction strength $UN$ we used the values 2.38, 2.23,
2.15, 1.75 for $\chi_{0-}$ being $49\%, 46\%, 43\%$ and $36\%$,
which fall within a few percent of those determined for the
corresponding experimental traces, however reduced by a common
scaling factor 1.89. The need for this factor is not surprising,
because in our atom number measurements up to a factor two
uncertainty should be expected for the absolute values, while
relative values are on the few percent level.

\begin{figure}
\includegraphics[width=8cm]{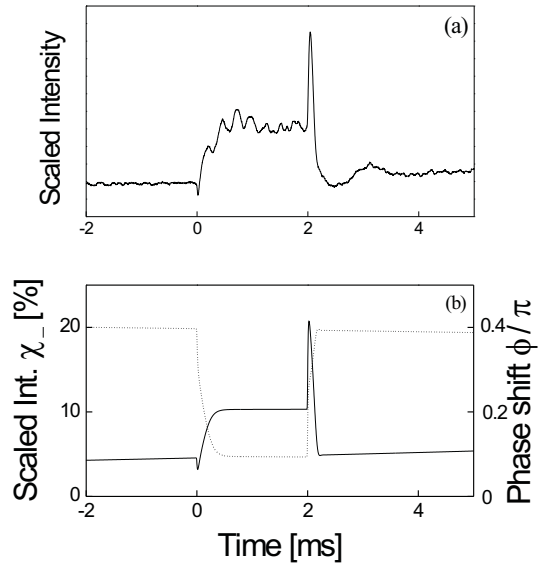}
\caption{ \label{jump} a) Observed intensity of the unlocked mode
plotted versus time. During the time interval 0\,ms$<$t$<2$\,ms
the total power coupled to the cavity is reduced by a factor of 2.
b) shows a numerical simulation based on the adiabatic model. The
intensity $\chi_-$ (solid line) and the corresponding phase $\phi$
(dotted line) are shown.}
\end{figure}

In the strong coupling regime our system shows an anomalous
response to a change of the total intensity $I_0$ coupled to the
cavity while keeping the same pumping asymmetry. This is shown in
Fig.~\ref{jump}a) for $\chi_{0-}=45\%$. In this plot we have
reduced $I_0$ by a factor two at $t=0$\,ms and doubled it again at
$t=2$\,ms. As a first immediate reaction to the reduction $\chi_-$
rapidly drops on a time scale given by the cavity decay time
${\gamma_{c}}^{-1}$, as might be expected. However, this is
counteracted by an approach of an increased steady state value on
a slower time scale. When the old power level is reestablished,
after a transient increase, $\chi_-$ drops back to nearly its
original value. A numerical simulation based on our adiabatic
model of Sec.~\ref{adiabatic} reproduces the experimental findings
surprisingly well except the an unexplained additional ripple at
about 3\,kHz, which amounts to about six times the radial
vibrational frequency. The calculations also show that during the
drop of $I_0$ the phase shift of the unlocked mode with respect to
the locked mode is reduced. Therefore the effective wavelength of
the unlocked mode is shifted closer to resonance and the intensity
increases.

\begin{figure}
\includegraphics[width=7cm]{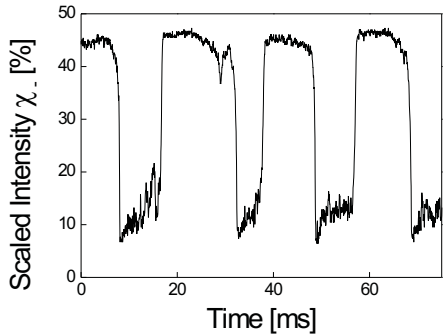}
\caption{ \label{switch} Bistable switching of $\chi_-$ during MOT
loading for $\chi_{0-}=49\%$}
\end{figure}

The bistable character of the atom--cavity system can also be
observed at work during MOT loading, yielding plots as that shown
in Fig. \ref{switch}, where $\chi_-$ switches between two steady
state solutions ($\chi_{0-}=49\%$). Starting at t=0 at the high
intensity level and hence with a deep lattice with comparably low
atom number $N$, the MOT tends to increase $N$ until the
interaction parameter exceeds the critical value and a jump
occurs. Now the intensity is low and hence the lattice depth,
while the temperature remains the same. This yields a decrease of
the loading rate and thus a reduction of $N$ until the system
jumps back to the previous state.

\begin{figure}
\includegraphics[width=8cm]{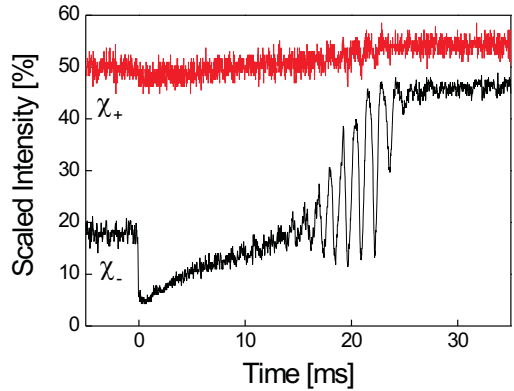}
\caption{The intra--cavity intensities $\chi_-$ and $\chi_+$ of the
unlocked mode (lower trace, recorded in transmission) and the
locked mode (upper trace, obtained from the reflected light) are
plotted versus time for $\chi_{0-}$=$45\%$.} \label{stable}
\end{figure}

In order to control the performance of the locking the light
reflected from the cavity originating from the locked travelling
wave mode is monitored together with the light of the unlocked
mode transmitted through one of the high reflectors . This allows
to verify, that despite of the rapid changes in time observed for
$\chi_-$ and $\phi$, the intra--cavity intensity of the
counterpropagating locked mode remains well behaved. This is
illustrated in Fig.~\ref{stable} for $\chi_{0-}$=$45\%$. While in
the lower trace rapid time evolution is observed the upper trace
remains nicely constant. The residual structure seen in the upper
trace is explained by imperfect separation of the contributions
from the two modes due to limited quality of the polarization
optics used.

\section{Non--adiabatic Motion}
\label{non--adiabatic}

\begin{figure}
\includegraphics[width=7cm]{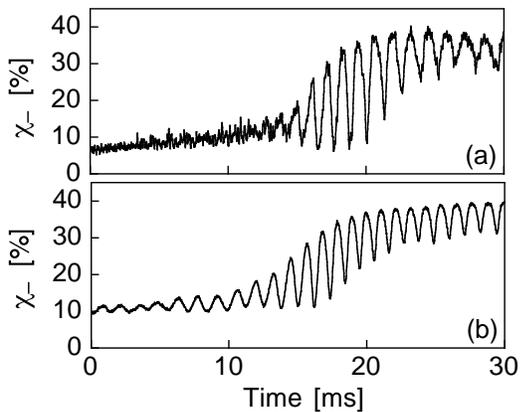}
\caption{ \label{simnon} a) Experimental observation of radial
breathing oscillations. (b) Simulation of the oscillations of (a)
by means of solving the complete set of equations of motion for
hundred atoms with an upscaled interaction strength.}
\end{figure}

The adiabatic model excellently explains the experimental findings
regarding the intra--cavity intensity except for the pronounced
oscillation feature observed in the $\chi_{0-}=43\%$--trace of
Fig.~\ref{intensities}. As has been discussed at the end of
Sec.~\ref{model} the assumption of adiabaticity is not well
justified for the radial degrees of the atomic motion. We came to
the conclusion that for fast changes of the lattice well depth
breathing oscillations should be excited which would produce
corresponding oscillations in the intra--cavity intensity, if the
system is operated near the frontier between the stable and the
bistable regime, i.e., in the $43\%$ trace of
Fig.~\ref{bistability}. In fact this oscillation is observed in
Fig.~\ref{intensities}. An expanded version is shown in Fig.~\ref{simnon}a).

\begin{figure}
\includegraphics[width=7cm]{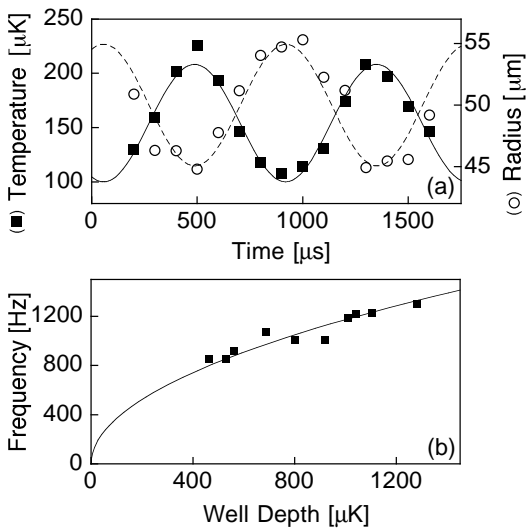}
\caption{ \label{oscillation} (a) Oscillation of the atomic
position spread (circles) and momentum spread (rectangles). The
solid and dashed lines are trigonometric fits with $\pi$ phase
delay. (b) Oscillation frequency plotted versus the well depth.
The solid line shows the expected square root dependence.}
\end{figure}

As an experimental test of our interpretation in terms of radial
breathing oscillations we have measured the momentum and position
spread of the atomic ensemble in the radial direction during the
observed intensity oscillations, finding the behavior shown in
Fig.~\ref{oscillation}a). The black rectangles show the radial
momentum spread of the atomic sample determined by
time--of--flight measurements, while the open circles show the
radial spread directly observed via in--situ images of the atoms
in the lattice. The solid and dashed lines are trigonometric fits
with $\pi$ phase delay, which confirm the expected anti--cyclic
behavior. In Fig.~\ref{oscillation}b) we plot the frequency of the
observed intensity oscillations versus the potential well depth
confirming the expected square--root dependence.

In order to include non--adiabatic aspects in our theoretical
description we use the full set of 6N+2 equations of motion of
Eq.~(\ref{fieldscaled}) and Eq.~(\ref{coordinates}) (i.e., 6N equations for
the atomic positions and momenta and two equations for the amplitude and the
phase of the unlocked intra--cavity mode). Since these equations
are coupled and non--linear, the simulation of all $10^6$ atoms is
beyond our computational capacities. Therefore, in our calculations
we reduce the number of atoms to one hundred and work with an
increased light shift per photon, such that the interaction
strength acquires values which compare to the experiments. The
artificially increased light shift per photon comes along with a
correspondingly increased light shift acting on each atom. Hence,
in order to maintain the potential well depth at the level used in
the experiments, we work with correspondingly decreased incoupled
intensities.

The full model for one hundred atoms very accurately reproduces
our calculations based on the adiabatic model shown in Fig.
\ref{simulation} except for the fact that in the vicinity of
$\chi_{0-}\approx 43\%$ additional oscillations arise. This
confirms once again that the assumption of adiabaticity of Sec.
\ref{adiabatic} is well justified for the axial degrees of
freedom. We can quantitatively reproduce the frequency of the
squeezing oscillation as is shown in Fig. \ref{simnon} b). The
experimental data are taken for $UN\approx 3.5$ and an axial
vibrational frequency of 500\,kHz, whereas for the theoretical
curve $UN=2.0$ and 550\,kHz was used. Applying the scaling factor
1.89 compensating for our systematic overestimation of the atom
number similarly as in Sec. \ref{non--linear} the measured
interaction strength coincides with the theoretical value within
$10\%$. The discrepancy in the vibrational frequencies lies within
our observation uncertainty of ca. $10\%$. The non--adiabatic
simulations of the system also reproduce the small frequency
decrease in time observed in Fig. \ref{simnon}a).

\section{Conclusions}
We conclude that optical lattices formed inside high finesse
cavities open up an interesting new regime characterized by
collective interactions significantly contributing to the atomic
dynamics. This regime can be experimentally realized even far from
an atomic resonance such that the interaction is entirely of
dispersive nature. Although the combination of a high finesse with
a large mode volume yields a small cavity resonance bandwidth, the
experimental challenge of preventing intra--cavity intensity
fluctuations can be handled. In the ring cavity studied here,
specific parameter ranges are identified which allow to operate a
stable lattice independent of the number of trapped atoms. Other
regimes are found where dispersive optical bistability accompanied
by self--induced breathing oscillations occur. The spatial phase
of the lattice is not pinned by the phases of the incoupled laser
beams, but rather determined by the strength of the atom--cavity
interaction. The bistable behavior arises for asymmetric pumping
and can be understood in terms of an adiabatic approximation for
the atomic motion, whereas the general equations of motion must be
considered to model the observed breathing oscillations. In this
article we have only discussed selected aspects of the system
dynamics. Other interesting phenomena could be studied, as for
example collective atomic recoil lasing (CARL) \cite{Bon:94},
which has been recently observed for unidirectional pumping of the
ring cavity \cite{Kru2:03}. The strong coupling regime might also
be utilized for implementing novel laser cooling schemes, which
rely on cavity--tailored coherent scattering, for example as
described in Ref.~\cite{Els:03}. Such schemes are highly desirable
since they promise to extend laser cooling to new species and to
operate in a density regime not yet accessible.

\begin{acknowledgments}
This work has been supported by Deutsche
For\-schungs\-gemeinschaft (DFG) under contract number $He2334/3-2$.
\end{acknowledgments}

\end{document}